\documentclass{aa}
\pdfoutput=1
\usepackage{graphicx}
\usepackage[varg]{txfonts}
\usepackage{amsmath}
\usepackage{verbatim}
\usepackage{latexsym}
\usepackage{multirow}

\begin{document}

\title{Data compression for local correlation tracking of solar granulation}

\author{B. L\"optien\inst{1,2}
\and A.~C. Birch\inst{2}
\and T.~L. Duvall Jr.\inst{2}
\and L. Gizon\inst{2,1}
\and J. Schou\inst{2}}

\institute{Institut f\"ur Astrophysik, Georg-August Universit\"at G\"ottingen, 37077 G\"ottingen, Germany
\and Max-Planck-Institut f\"ur Sonnensystemforschung, Justus-von-Liebig-Weg 3, 37077 G\"ottingen, Germany}

\date{Received <date> /
Accepted <date>}

\abstract {Several upcoming and proposed space missions, such as {\it Solar Orbiter}, will be limited in telemetry and thus  require data compression.}
{We test the impact of data compression on local correlation tracking (LCT) of time-series of continuum intensity images. We evaluate the effect of several lossy compression methods (quantization, JPEG compression, and a reduced number of continuum images) on measurements of solar differential rotation with LCT.}
{We apply the different compression methods to tracked and remapped continuum intensity maps obtained by the {\it Helioseismic and Magnetic Imager} (HMI) onboard the {\it Solar Dynamics Observatory}. We derive 2D vector velocities using the local correlation tracking code FLCT and determine the additional bias and noise introduced by compression to differential rotation.}
{We find that probing differential rotation with LCT is very robust to lossy data compression when using quantization. Our results are severely affected by systematic errors of the LCT method and the HMI instrument. The sensitivity of LCT to systematic errors is a concern for {\it Solar Orbiter}.}
{}
\keywords{Sun: helioseismology - Methods: data analysis}

\maketitle

\section{Introduction}
Local correlation tracking of granulation~\citep[LCT,][]{1988ApJ...333..427N} is an important method for measuring flows in the photosphere such as supergranulation or large-scale flows. It will play a significant role in upcoming and proposed space missions, such as {\it Solar Orbiter}~\citep{Yellowbook,Redbook,2014SSRv..tmp...31L}. Unfortunately, the data rate of {\it Solar Orbiter} will be limited and data compression will be required.

In this paper, we evaluate how continuum intensity images can be compressed with as little influence on the derived LCT flow maps as possible. We use data provided by the {\it Helioseismic and Magnetic Imager}~\citep[HMI,][]{2012SoPh..275..229S} and the local correlation tracking code FLCT~\citep{2004ApJ...610.1148W,2008ASPC..383..373F}. We focus on measuring differential rotation, one of the key science objectives of {\it Solar Orbiter}. We combine two different approaches for compressing the data. In the first case,	the individual continuum images are compressed using quantization or JPEG compression, similar to what we tried in time-distance helioseismology~\citep{2014A&A...571A..42L}. In the second case, the time lag between the intensity images used by the LCT code is increased, reducing the number of images. These two approaches allow us to obtain first estimates of the performance of these compression methods.

\section{Data and methods}

\subsection{Input data}\label{sect:data}
The analysis is based on 120 days of tracked and remapped HMI continuum intensity images from 1 May to 28 August 2010~\citep[provided by][]{2015A&A...581A..67L}. The cubes have a size of $178\times 178$~Mm ($512\times 512$~pixels with $0.348$~Mm pixel size) and are centered around six latitudes along the central meridian ($0^\circ$,$10^\circ$,$20^\circ$,$40^\circ$,$60^\circ$ north and $60^\circ$ south) and at $60^\circ$ longitude east and west at the equator. The cubes have a cadence of 45~s and are tracked for 24~h each using the Mt. Wilson 1982/84 differential rotation rate~\citep{1984SoPh...94...13S} corresponding to the central latitude of each cube.

\subsection{The FLCT code}
The FLCT code~\citep[Fourier Local Correlation Tracking,][]{2004ApJ...610.1148W,2008ASPC..383..373F} computes 2D maps of horizontal flows by tracking the motion of small features on the Sun, in our case granulation in HMI continuum images. The FLCT code has mostly been used for studying the evolution of magnetic fields in the photosphere~\citep[e.g.,][]{2010ApJ...723.1651K,2012ApJ...758L..38O}. 

The code measures flows by cross-correlating pairs of continuum intensity images separated in time. First, the two images are split into subimages which are apodized with a Gaussian with a width $\sigma$. Then, the 2D cross-covariances in $x$ (east-west direction, increasing westwards) and $y$ (north-south direction, increasing northwards) between the subimages are computed. The position of the maximum of the cross-covariance function is the shift between the two subimages and can be used to determine the flow velocities. The parameter $\sigma$ defines the resolution of the resulting flow maps. Here we use $\sigma=6$~pixels ($\sim 2.1$~Mm). The output of the FLCT code is a flow map for every pair of consecutive images in the input data. Running FLCT on the data described in Section~\ref{sect:data} results in a 24~h long time-series of LCT flow maps with a cadence of 45 s. In the following, we use averages of the flow maps over 24~h.

\subsection{Compression methods} \label{sect:comp}
\subsubsection{Quantization}
Quantization compresses the data by dividing them by a scaling factor and rounding to the nearest integer. This reduces the number $n$ of possible values that the intensity can assume and thus the number of bits per pixel needed for storing the data. The lower $n$, the lower the precision of the data and the smaller the file size.

We apply quantization to the data using fixed values of $n$ (between 2 and 256) by varying the scaling factor. Before applying the quantization, we subtract a mean image from the data in order to remove limb darkening. Limb darkening has the largest contribution to the intensity contrast at high latitudes. Without removing the mean image, quantization would predominantly sample limb darkening but not granulation.

As described in~\citet{2014A&A...571A..42L}, we reduce the file size further by running lossless Huffman encoding~\citep{Huffman} on the quantized data. Surrounding pixels in space and time are highly correlated. We use this to increase the efficiency of the compression.

\subsubsection{JPEG compression}
JPEG (Joint Photographic Experts Group) compression~\citep{Wallace1992} reduces the file size of images by removing information about small spatial scales. JPEG compression divides each continuum image into $8\times 8$~pixel blocks and applies a discrete cosine transform (DCT) to every block. The coefficients of the transformation are then truncated depending on a quality factor {\it q} (between 0 and 100, with a lower factor causing a higher compression). After truncating, the coefficients of the DCT are compressed using lossless Huffman compression.

Like for quantization, we subtract the mean image before applying JPEG compression to the data in order to remove limb darkening. We test different values of the quality factor $q$ (between 10 and 100).

\subsubsection{Sampling rate}
Obtaining LCT flow maps every 45~s is not very efficient in reducing the noise level. The main source of noise in LCT arises from the proper motion of granules. Granules have a lifetime of about ten minutes and so, their proper motion in LCT flow maps that are separated by 45~s is highly correlated. Increasing the time lag $\Delta T$ between consecutive flow maps but keeping a cadence of 45~s between consecutive intensity images only slightly increases the noise in the final flow map averaged over 24~h, as long as $\Delta T$ is smaller than the granulation lifetime. Long time lags, however, increase the granulation noise in the final time-averaged flow map significantly.

This method does not compress the individual intensity maps but reduces the number of intensity maps required for running the LCT code. Instead of using all the images provided by HMI (separated by 45~s), we use pairs of images with the images within the pairs being separated by 45~s, but consecutive pairs by a longer time $\Delta T$ (between 45~s and 180~min). We also test the combination of this method with compressing the individual intensity images by using quantization or JPEG compression.

The time lag $\Delta T$ slightly decreases the efficiency of the Huffman compression, because consecutive intensity images are not highly correlated anymore.

\subsection{Calibration of LCT velocities}\label{sect:calibration}
\begin{figure}
\centering
\includegraphics[width=\columnwidth]{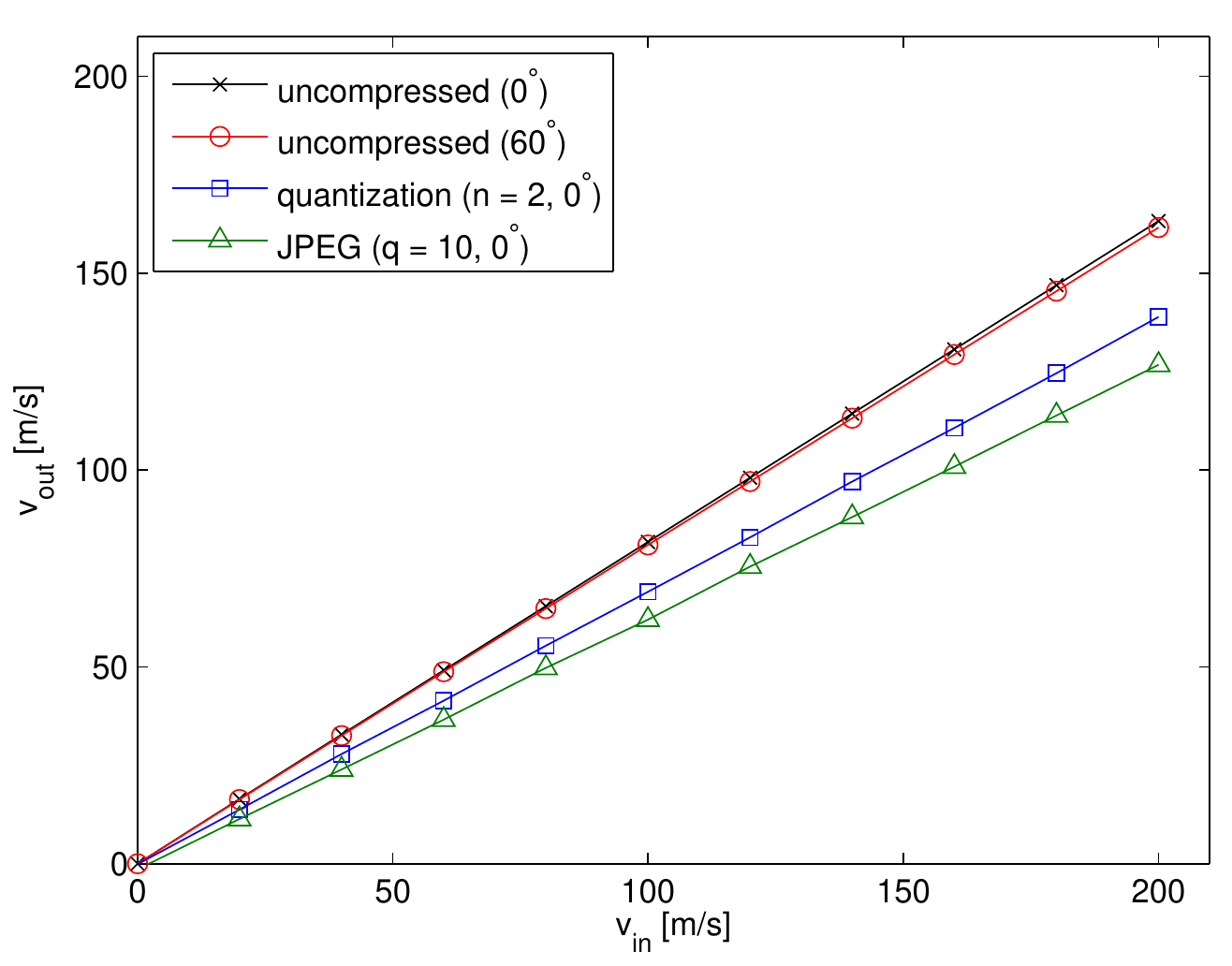}
\caption{Calibration curves for LCT, {\it black crosses:} uncompressed data at $0^\circ$ latitude, {\it red circles:} uncompressed data at $60^\circ$ latitude, {\it blue squares:} quantized data ($n = 2$) at $0^\circ$ latitude, {\it green triangles:} JPEG compression ($q = 2$) at $0^\circ$ latitude. The error bars are smaller than the symbol size. The amplitudes of the velocities generated by LCT are in general lower than the actual velocities on the Sun, especially for compressed data. We correct for this effect by generating calibration data with a constant flow in the $x$-direction. Three days of data are used for each data point. For more details about this method see the text.}
\label{fig:calibration}
\end{figure}

In general, LCT tends to underestimate flow velocities~\citep[see e.g.,][]{2007SoPh..241...27S,2013A&A...555A.136V}. For measuring rotation, the amplitudes of the velocities are important. These are corrected by generating calibration curves for the LCT velocities. Our calibration data consist of HMI continuum intensity data, to which we add a constant flow in the $x$-direction by shifting the individual images using Fourier interpolation. Most the effects that could potentially cause the low amplitudes of the LCT velocities are also present in the calibration data. The flows determined from these data with LCT are a superposition of the known input velocity and the (not known) solar flows (supergranulation, rotation). In order to determine the contribution of the calibration signal to the LCT velocities, a calibration curve must be generated by shifting the same datacube using different input velocities. The contribution from solar flows should be the same for all calibration velocities. The slope of 
the LCT velocities as a function of the input velocities can then be used to correct for the underestimation of velocities by LCT. Example calibration curves are shown in Figure~\ref{fig:calibration}. The slope of the curve depends slightly on the distance to disk center and is severely affected by compression. Quantization and JPEG reduce the sensitivity of LCT, a change of the time lag $\Delta T$ does not change the slope. This calibration method assumes that the sensitivity of LCT to the actual solar flows does not depend on the spatial scale.

\section{Results}
\subsection{Individual flow maps}
\begin{figure*}
\centering
\includegraphics[width=17cm]{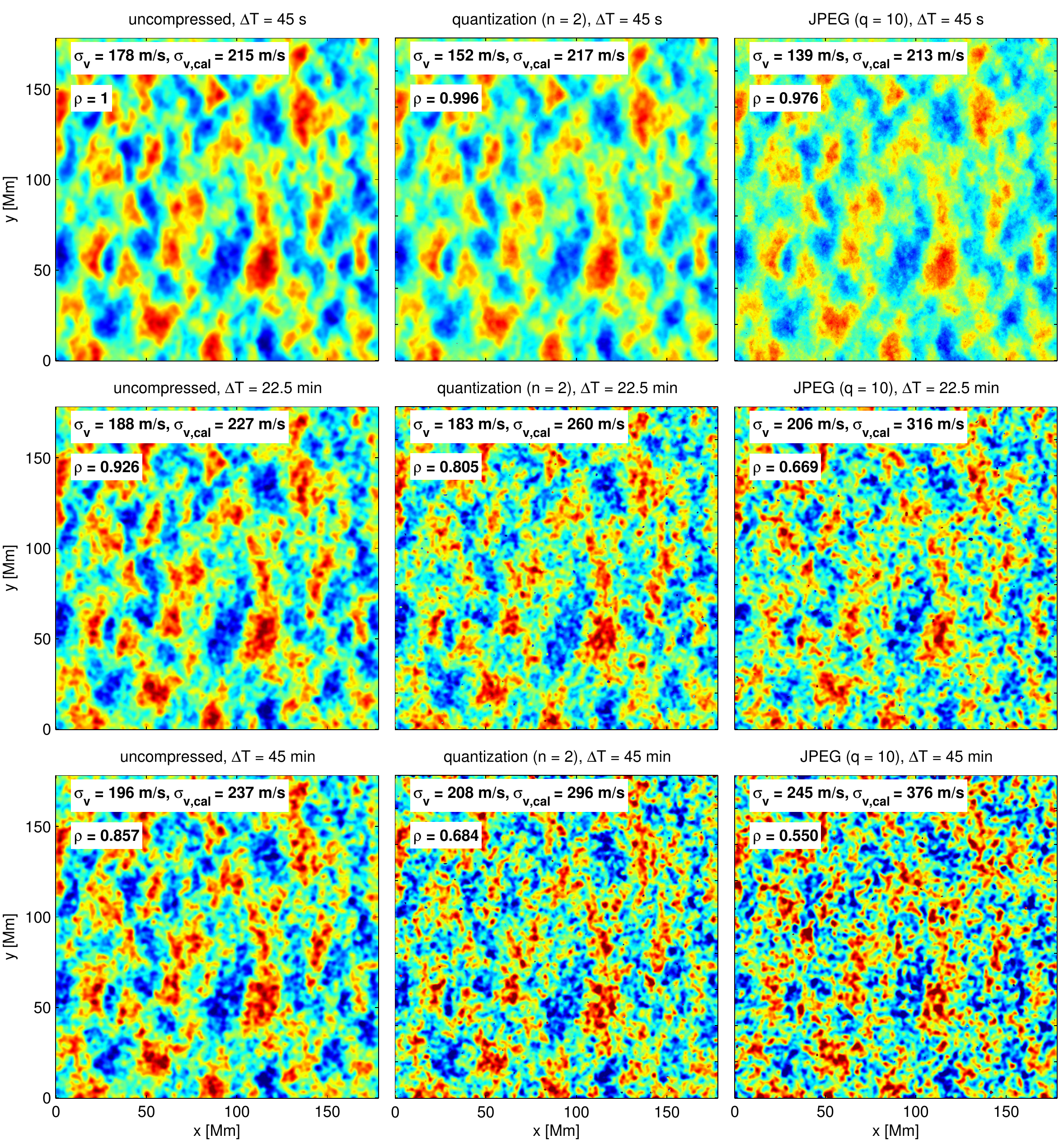}
\caption{Flows in the $x$-direction at disk center, as determined by local correlation tracking of granulation (averaged over 24 h). We compare a flow map for uncompressed data with flow maps for different compression methods. {\it Left column:} uncompressed data, {\it middle column:} quantization ({\it $n = 2$}), {\it right column:} JPEG (quality = 10). In the vertical direction, we vary the time lag $\Delta T$ between consecutive velocity maps used for averaging over 24~h ({\it top row:} $\Delta T = 45$~s, {\it middle row:} $\Delta T = 22.5$~min, {\it bottom row:} $\Delta T = 45$~min). All flow maps clearly show supergranulation. We use the same scaling of the colormap for all images, corresponding to the amplitudes of the velocities before applying the calibration factor. The numbers in the images give the standard deviation of the velocities before ($\sigma_v$) and after ($\sigma_{v,\rm cal}$) applying the calibration factor to the data and the correlation with the LCT maps computed from uncompressed 
data (
$\rho$). The flow maps computed from compressed data have a higher noise level than those for uncompressed data. Quantization and JPEG lead to noise 
in the cross-covariances used by the LCT code and, thus, results in noise that varies on a pixel-by-pixel scale. JPEG causes a higher noise level than quantization. A large time lag between consecutive velocity maps causes granulation noise to remain visible in the image.}
\label{fig:flow_maps}
\end{figure*}
\begin{figure}
\centering
\includegraphics[width=\columnwidth]{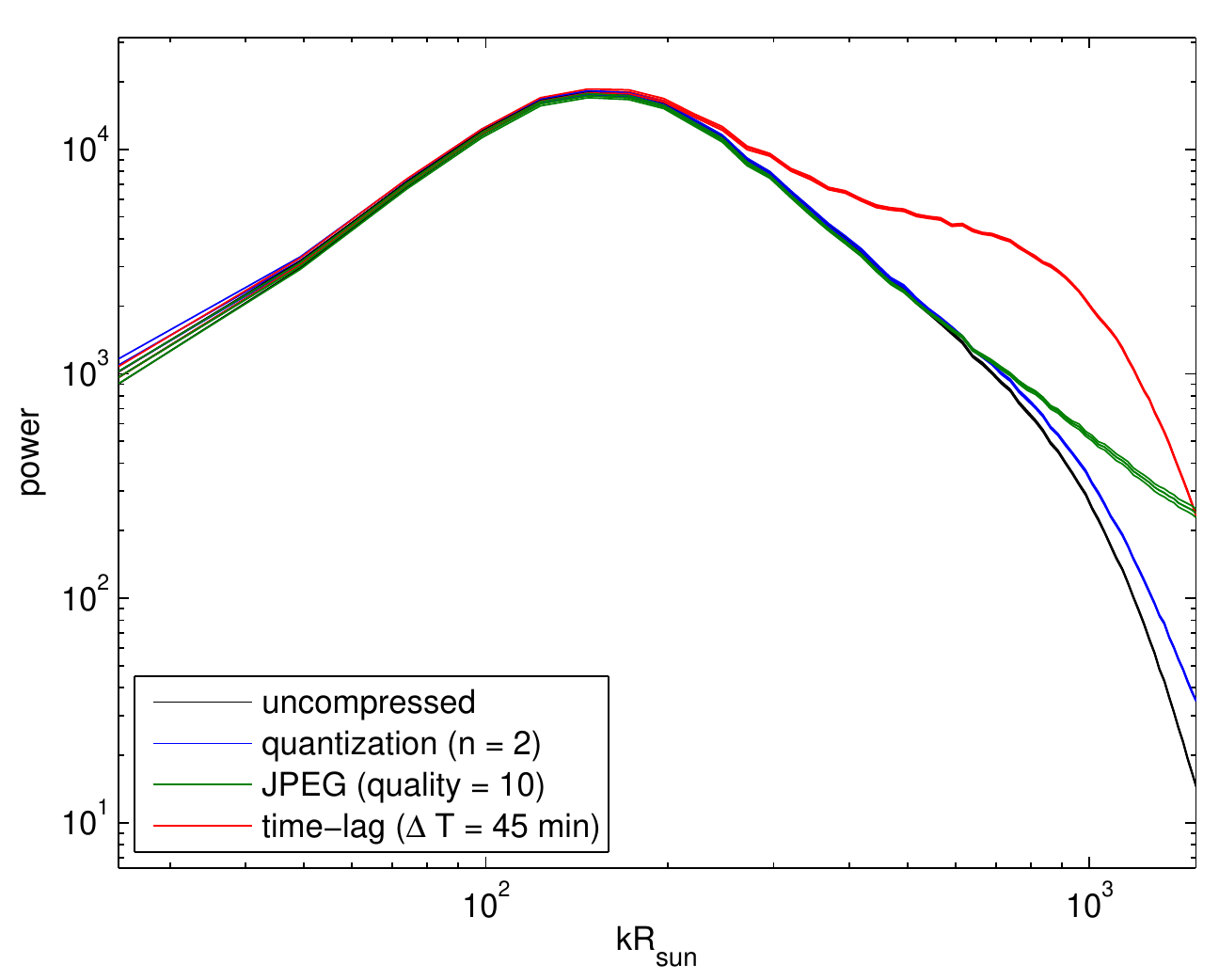}
\caption{Spatial power spectra of the flow divergence at disk center (averaged power for $T = 120$~days). {\it Black:} uncompressed data, {\it blue:} quantization ({\it n = 2}), {\it green:} JPEG (quality = 10), {\it red:} large time lag between consecutive velocity maps ($\Delta T = 45$~min). The thickness of the lines shows the $1\sigma$-scatter of the mean power. All curves exhibit a peak at $kR_\odot \approx 120$ resulting from supergranulation. At high wavenumbers the additional noise introduced by the compression is visible. A large time lag between consecutive velocity maps does not remove all the noise from granulation. Quantization and JPEG lead to noise on a pixel-by-pixel scale with JPEG being worse than quantization. At supergranulation scales, the influence of quantization and JPEG compression is negligible.}
\label{fig:power_spectra}
\end{figure}

Figure~\ref{fig:flow_maps} shows flow velocities in the $x$-direction determined by local correlation tracking, both for uncompressed data and for the various compression methods discussed in Section~\ref{sect:comp}. We study the performance of the compression for a broad range of parameters (intensity bins $n$, quality factor $q$, and time-lag $\Delta T$), but the results that we present in this Section and in the following ones are mostly for the highest compression factors because these allow us to easily identify the artifacts introduced by the various compression algorithms. In all flow maps, supergranulation is clearly visible, but the noise level is much higher for the compressed data. Quantization and JPEG compression cause noise in the cross-covariances computed by the FLCT code, leading to additional noise in the derived velocities on a pixel-by-pixel scale. When a large time lag $\Delta T$ between pairs of intensity images is used, the noise from the proper motion of the granules remains in the 
flow maps. Quantization and JPEG compression also reduce the amplitude of the velocity, but this can be compensated using the calibration curves discussed in Section~\ref{sect:calibration}. The increased noise level is also visible in spatial power spectra of the flow divergence (Figure~\ref{fig:power_spectra}), where it increases the power at high wavenumbers. However, at wavenumbers corresponding to the length scale of supergranules ($kR_\odot \approx 120$), the influence of quantization and JPEG compression is negligible.

This can also be seen when deriving the standard deviation of the flow maps and the correlation with the LCT flow maps computed from uncompressed data (see Table~\ref{tab:flow_maps}). In case of the nominal flow maps provided by the FLCT code, the compression has a large influence on the standard deviations and the correlation coefficients. However, if the FLCT flow maps (both for the uncompressed and for the compressed data) are smoothed by convolving them with a Gaussian (width $\sigma_G = 3$~Mm), the effect of the compression on the standard deviations and the correlation coefficients is much smaller.

The performance of JPEG compression is worse than quantization. It increases the noise level more than quantization and is also less correlated with the uncompressed case.

The amount of power of high wavenumbers depends on the apodization parameter $\sigma$. Higher values of $\sigma$ reduce the sensitivity of LCT to small spatial scales. However, the compression causes an excess in power at high wavenumbers for all values of $\sigma$ that we have tested ($\sigma$ = 3, 6, and 10 pixels).

\begin{table*}
\caption{Standard deviation of the flow maps for $v_x$ after applying the calibration factor and the correlation with the LCT flow maps computed from uncompressed data. We show results both for the nominal flow maps provided by the FLCT code (as shown in Figure~\ref{fig:flow_maps}) and for the same flow maps after convolving them (both the uncompressed and the compressed data) with a Gaussian with a width of $\sigma_G = 3$~Mm.}
\label{tab:flow_maps}
\centering
\begin{tabular}{p{3cm}llllll}
\hline\hline
compression method & time lag $\Delta T$ & \multicolumn{2}{c}{$\sigma$ [m/s]} & \multicolumn{2}{c}{corr. with uncompressed}\\
 & & without filtering & with filtering& without filtering & with filtering\\
\hline
uncompressed & 45 s& 215 & 159 & 1 & 1\\
 & 22.5 min & 227 & 158 & 0.926 & 0.983\\
 & 45 min & 237 & 155 & 0.857 & 0.964\\
quantization ($n = 2$) & 45 s& 217 & 159 & 0.996 & 0.999\\
 & 22.5 min & 260 & 162 & 0.805 & 0.956\\
 & 45 min & 296 & 164 & 0.684 & 0.912\\
JPEG ($q = 10$) & 45 s & 213 & 154 & 0.976 & 0.997\\
 & 22.5 min & 316 & 165 & 0.669 & 0.907\\
 & 45 min & 376 & 174 & 0.550 & 0.835\\
\hline
\end{tabular}
\end{table*}

\subsection{Shrinking-Sun effect}
\begin{figure*}
\centering
\includegraphics[width=17cm]{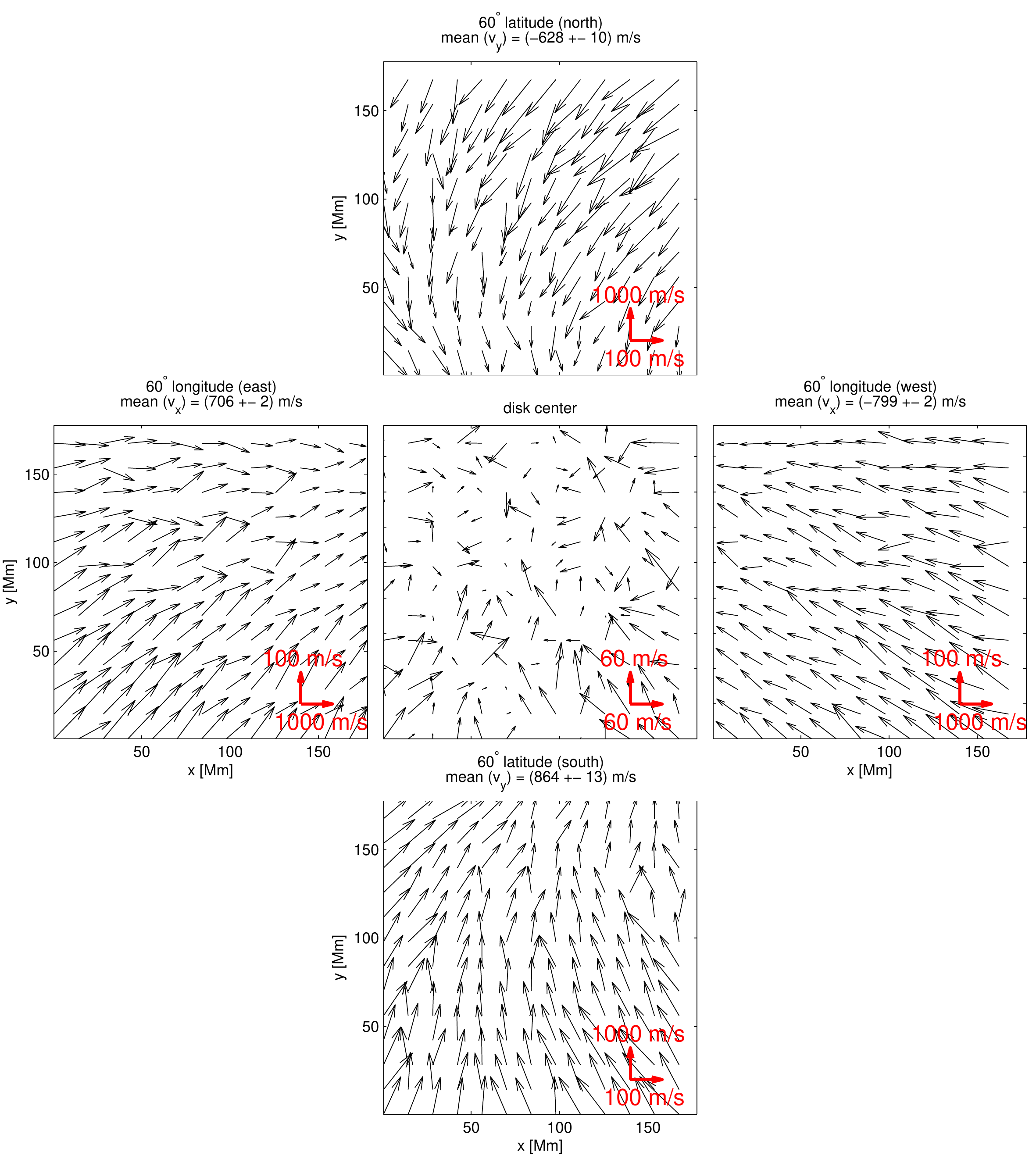}
\caption{Vector velocities derived by LCT averaged over 120 days for several positions on the disk. {\it Top:} velocities at $60^\circ$ latitude north at the central meridian, {\it bottom:} $60^\circ$ latitude south at the central meridian, {\it left:} $60^\circ$ longitude east at the equator, {\it center:} $0^\circ$ latitude and $0^\circ$ longitude, and {\it right:} $60^\circ$ longitude west at the equator. The scaling of the arrows (shown in the {\it bottom right} in the individual images) for the plot at disk center is different to the other ones. The range of the velocities varies significantly over the disk due to different magnitudes of the shrinking-Sun effect. We amplify the flows in the azimuthal direction ($v_x$ for $60^\circ$ latitude north and south and $v_y$ for $60^\circ$ longitude east and west) by a factor of ten when displaying the arrows. The mean values of the velocities in radial direction are shown above the individual images. The amplitude of the shrinking-Sun effect is different for 
$60^\circ$ latitude north and south. This is caused by the $B_0$-angle. It is mostly positive and increases from $-4^\circ$ to $+7^\circ$ over the course of the 120 days that we have studied. There is also a difference of $(93\pm 3)$~m/s in the amplitude of the shrinking-Sun effect between $60^\circ$ longitude east and west. The reason for this is not understood yet. At the equator, the velocities in the $y$-direction caused by the shrinking-Sun effect point predominantly northwards due to the mostly positive $B_0$-angle.}
\label{fig:flow_maps_quiver}
\end{figure*}
Flow maps derived by LCT are severely affected by the shrinking-Sun effect, an artifact in LCT velocities that looks like a flow pointing towards disk center~\citep{2004ESASP.559..556L}. As can be seen in Figure~\ref{fig:flow_maps_quiver}, it dominates the velocities derived from LCT. The amplitude of the velocities caused by this effect increases with increasing distance from disk center. This ``flow'' reaches up to $1.0$~km/s. The relative contribution of this effect to flows in the $x$- and the $y$-direction depends on the position on the disk. Along the central meridian, for example, the main component is in the $y$-direction.  Close to the eastern and western sides of the tracked regions, the flow component in the $x$-direction also becomes significant. At $60^\circ$ latitude north, this leads to a variation of $v_x$ of about 80~m/s in the east-west direction. In theory, the shrinking-Sun effect should not affect measurements of rotation: it should be antisymmetric around the central meridian and cancel 
out when averaging over longitude. However, due to the large amplitude of this effect, even a small deviation from perfect antisymmetry can result in a significant bias of the derived rotation rate. Figure~\ref{fig:flow_maps_quiver} indeed shows a difference of $(93\pm 3)$~m/s between the amplitudes of the velocities in the $x$-direction at $60^\circ$ longitude east and west of the central meridian. This difference is too large to be caused by rotation (difference between solar rotation and tracking rate). Unfortunately, it is not possible to determine if the shrinking-Sun effect in $v_x$ along the central meridian exhibits an east-west asymmetry as well. This velocity is a superposition of the shrinking-Sun effect and the residual signal from differential rotation (rotation minus tracking rate). Solar rotation is constant in the east-west direction, lifting any asymmetry that might be caused by the shrinking-Sun effect.

\subsection{Differential rotation}
\begin{figure*}
\centering
\begin{minipage}{8cm}
\includegraphics[width=\columnwidth]{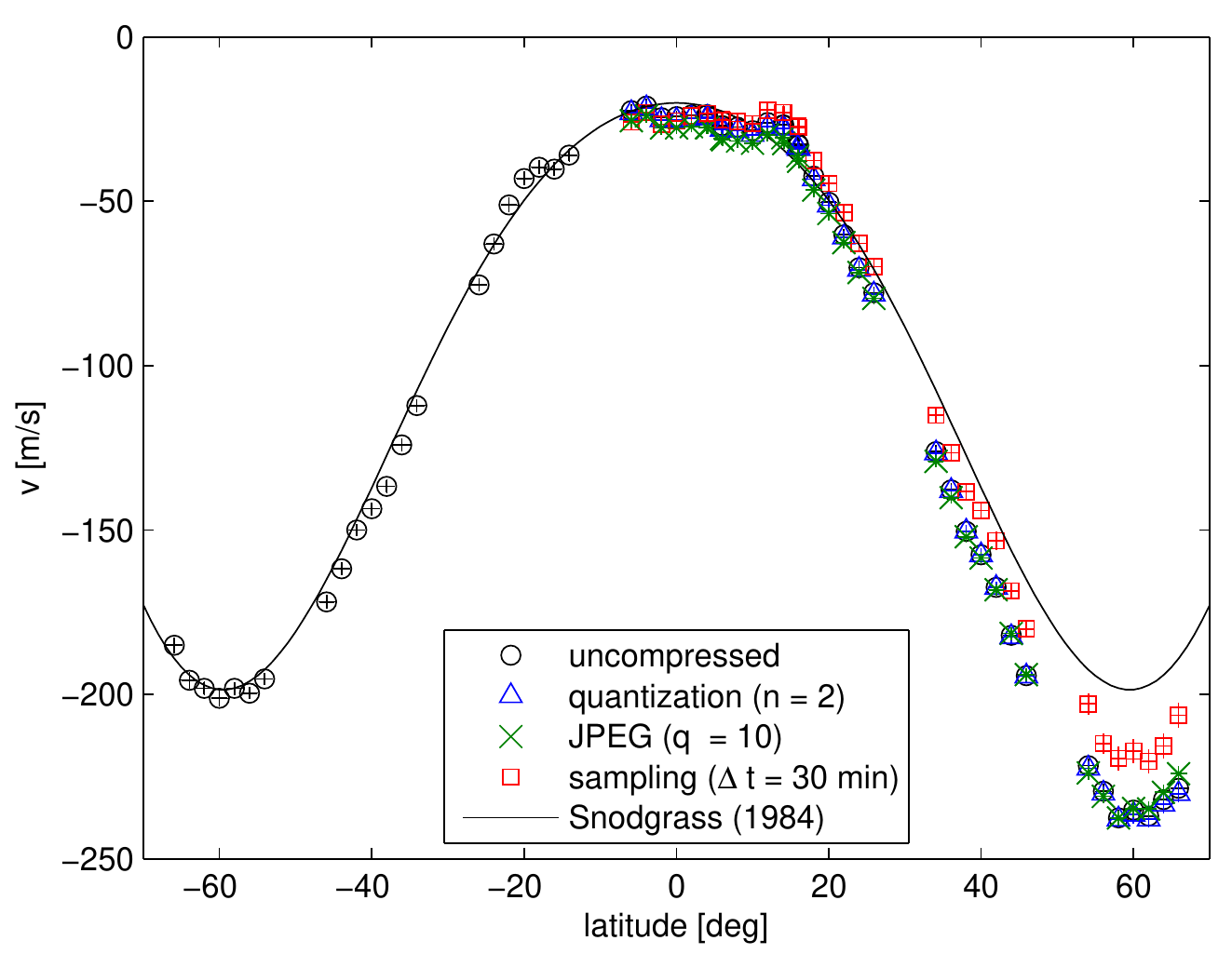}
\end{minipage}
\begin{minipage}{8cm}
\includegraphics[width=\columnwidth]{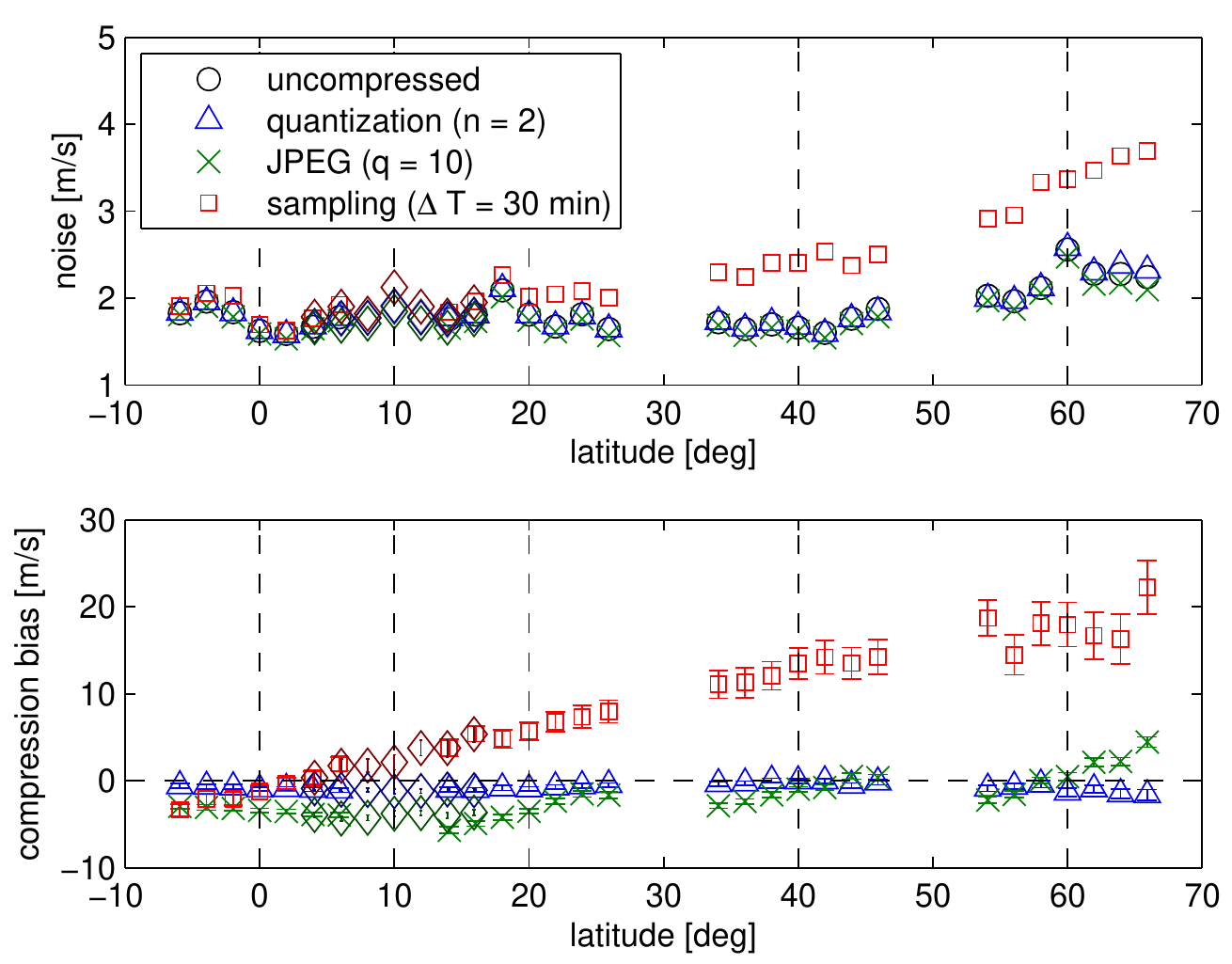}
\end{minipage}
\caption{{\it Left:} differential rotation relative to solid body rotation determined using LCT on uncompressed and compressed data ($T = 120$~days). {\it Black circles:} uncompressed data, {\it blue triangles:} quantization (n = 2), {\it green crosses:} JPEG (quality = 10), {\it red squares:} large time lag between consecutive velocity maps ($\Delta T = 30$~min). The solid curve is the~\citet{1984SoPh...94...13S} rotation rate which is used by the tracking code. In order to save computation time, we only computed the rotation rate for positive latitudes with compressed data. In the northern hemisphere, there is a large (up to $\sim 34$~m/s) offset between the~\citet{1984SoPh...94...13S} rate and the velocities derived using LCT. Its origin is unknown. The error bars are smaller than the symbol size. {\it Right:} noise level ({\it top}) and compression bias (difference compressed minus uncompressed data, {\it bottom}) as a function of latitude for the differential rotation curve shown on the {\it left}. The 
vertical dashed lines show the centers of the tracked and remapped intensity images. The {\it diamonds} show the noise and bias derived from intensity data centered at $10^\circ$ latitude, overlapping with the data centered at $0^\circ$ and $20^\circ$ latitude. The time lag method suffers from a large compression bias that increases linearly with latitude. Quantization and JPEG are in good agreement with the uncompressed case.}
\label{fig:rotation}
\end{figure*}

Figure~\ref{fig:rotation} shows differential rotation derived from LCT data. The rotation rate from~\citet{1984SoPh...94...13S} is included for comparison. There are big differences between the rotation rate between the northern and southern hemispheres. At $60^\circ$ latitude, the southern hemisphere rotates faster by $(34\pm 4)$~m/s. In the southern hemisphere, our results are in agreement with the~\citet{1984SoPh...94...13S} rate, but in the northern hemisphere, there is an offset which increases with latitude. Of course, the Sun could have been rotating faster in the southern hemisphere than in the northern one for the studied time-period, but this difference could also be related to the shrinking-Sun effect (it is not perfectly antisymmetric along the equator, see Figure~\ref{fig:flow_maps_quiver}). The noise level is almost constant with latitude. Supergranulation is the main source of noise when probing differential rotation with LCT. The flow maps all have the same size ($178 \times 178$~Mm) and so, 
they contain the same number of supergranules and give a similar noise level.

Figure~\ref{fig:rotation} also demonstrates how the various compression methods affect measurements of differential rotation. In order to save computation time we only computed the rotation rate for positive latitudes from the compressed data. The time lag method exhibits a large bias that increases linearly with latitude. Quantization and JPEG compression are in good agreement with the uncompressed case. We also include data centered at $10^\circ$ latitude, overlapping with the intensity images centered at $0^\circ$ and $20^\circ$. The overlapping parts are in good agreement, suggesting that measuring rotation and compression are not affected by projecting the intensity images on a cartesian coordinate system.

\subsection{HMI $P$-angle oscillations}
The large bias of the time lag method is caused by oscillations of the $P$-angle of the HMI instrument that, as far as we know, have not been reported before. The roll angle of the SDO spacecraft relative to the rotation axis of the Sun is given by the keyword {\it CROTA2} and is saved for each individual filtergram obtained by HMI ($3.75$~s cadence). This keyword shows oscillations with a fundamental period of $27.8$~mHz and several harmonics. These oscillations are also present in the CROTA2 keyword of the 45~s cadence computed continuum intensity images. Due to aliasing the dominant frequency appears at $5.56$~mHz and two harmonics are at $2.78$ and $8.33$~mHz ($\sim 10^{-4}$~ deg rms in total). The origin of these oscillations is currently unclear. Such a periodic shift of the intensity images affects, of course, the derived LCT velocities. For data along the central meridian the velocities in the $x$-direction and for the cubes along the equator in the velocities in the $y$-direction exhibit 
oscillations at the same frequencies as the {\it CROTA2} keyword. The amplitude of the oscillations in the LCT data can be estimated from a simple model of a disk rotating around its center with an angular velocity given by the changes of the {\it CROTA2} keyword. However, the amplitudes in the power spectrum predicted by this simple model do not match exactly with the observations in LCT (about 10\% difference, after applying the calibration described in Section~\ref{sect:calibration}). In theory, these oscillations should be removed by the tracking code. This is not the case here, indicating that the {\it CROTA2} keyword does not describe the $P$-angle oscillations correctly.

\begin{figure}
\centering
\includegraphics[width=\columnwidth]{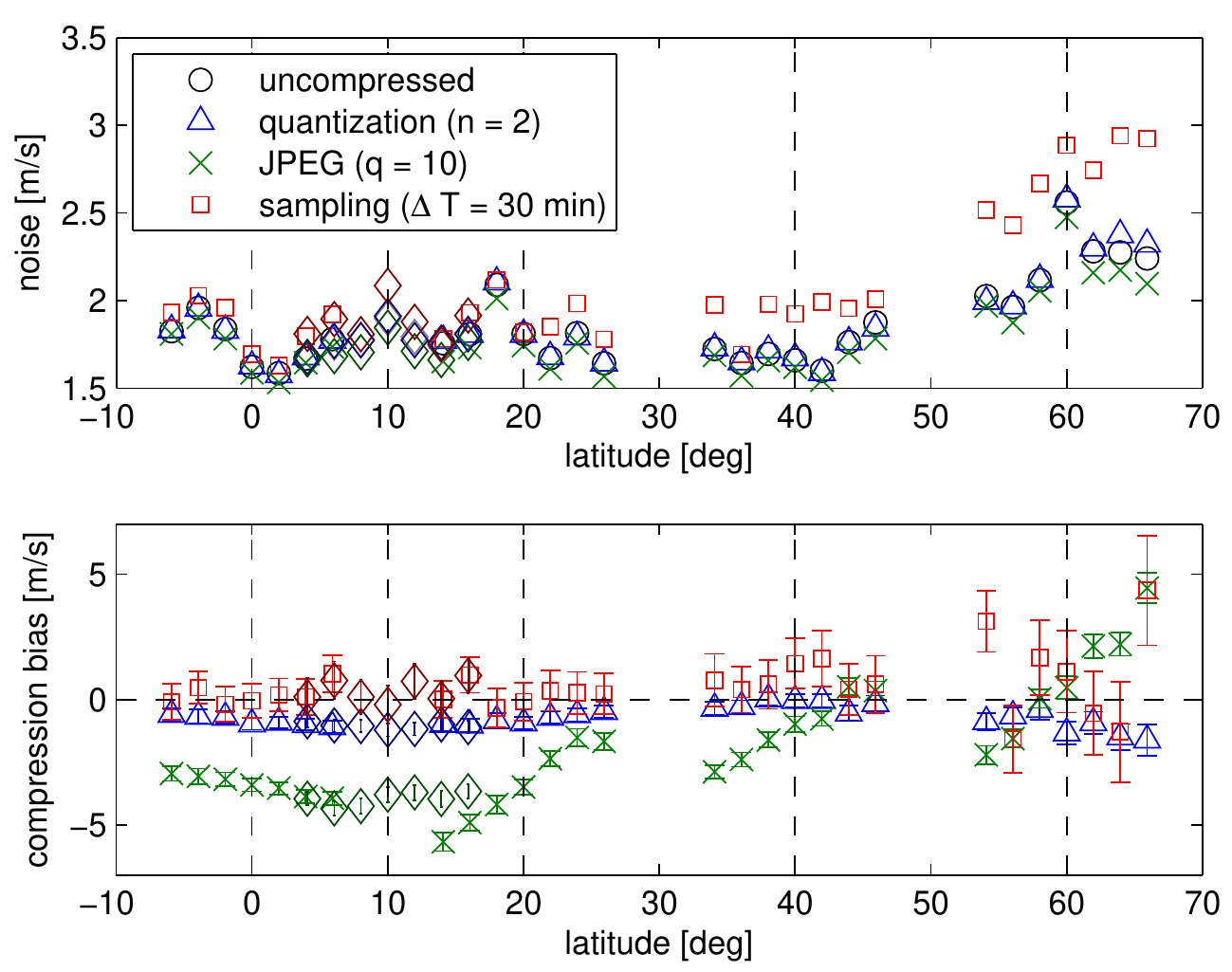}
\caption{Same as the right part of Figure~\ref{fig:rotation}, but with a simple correction method being applied to the time lag method that is based on the {\it CROTA2} keyword of the HMI data. More details are described in the text. This correction removes most of the bias and slightly reduces the noise level. The remaining bias of the time lag method is comparable to the noise level. The time lag method suffers from a higher noise level than the uncompressed data, since granulation noise is still present in the data. Quantization and JPEG compression have almost no influence on the noise level. The compression bias for quantization is comparable with the noise level. JPEG compression exhibits a strong bias that varies with latitude. Note that the ranges of the $y$-axes are different to those in Figure~\ref{fig:rotation}.}
\label{fig:noise_bias_corr}
\end{figure}

The oscillations have a short period and mostly cancel out when averaging a sufficient amount of data. However, if the time lag $\Delta T$ between consecutive velocity images is in resonance with these oscillations (as it is the case for $\Delta T = 30$~min, corresponding to a frequency of $0.556$~mHz), this effect can lead to an offset of the order of up to 20~m/s in the flow maps averaged over 24~h (as observed in Figure~\ref{fig:rotation}). In Figure~\ref{fig:noise_bias_corr}, we tried to remove these oscillations by using the simple model described above. The remaining bias is comparable to the noise level. There might still be some residual signal, though, because the amplitude of the effect predicted from the {\it CROTA2} keyword does not match exactly with that from the LCT velocities (as discussed above). In the remainder of this paper, we only show results after applying this correction.

\subsection{Influence of compression on differential rotation}
Compression affects LCT in several ways. It could potentially change the noise level, systematic errors, and also the signal from solar rotation itself. Determining the impact on the noise level is straightforward, but it is hard to measure the impact on the expectation value of the rotation velocities. It is only possible to determine the compression bias, defined as the difference compressed minus uncompressed. This compression bias is also affected by changes of the noise or systematic errors. For all compression methods, the expectation value of the bias does not depend on the observing time. Noise and compression bias for differential rotation are shown in Figure~\ref{fig:noise_bias_corr}.

The time lag method suffers mostly from the oscillations of the $P$-angle of HMI (as discussed in the previous section). In addition, the noise level is increased significantly due to the granulation noise. This probably also causes the remaining bias of the time lag method (after correcting for the $P$-angle oscillations). Quantization and JPEG compression have almost no influence on the noise level. The noise in the individual flow maps is on very small spatial scales and is removed almost entirely when averaging over longitude. The rotation velocities computed from these data are also strongly correlated with the uncompressed data, indicating that the bias is not caused by changes of the noise. In most cases, the compression bias introduced by quantization and JPEG compression is negative. Quantization and JPEG compression enhance the shrinking-Sun effect. If the deviation of the uncompressed case from the~\citet{1984SoPh...94...13S} rate is really caused by the shrinking-Sun effect, this offset will be 
amplified by compression, leading to a negative compression bias. Some of the bias of the JPEG method seems to be originating from inaccuracies of the calibration. The compression bias at around $15^\circ$ latitude is different for the data centered at $10^\circ$ and $20^\circ$ latitude. In addition, the bias for JPEG compression exhibits a linear trend within the individual datacubes.

\begin{figure*}
\centering
\includegraphics[width=17cm]{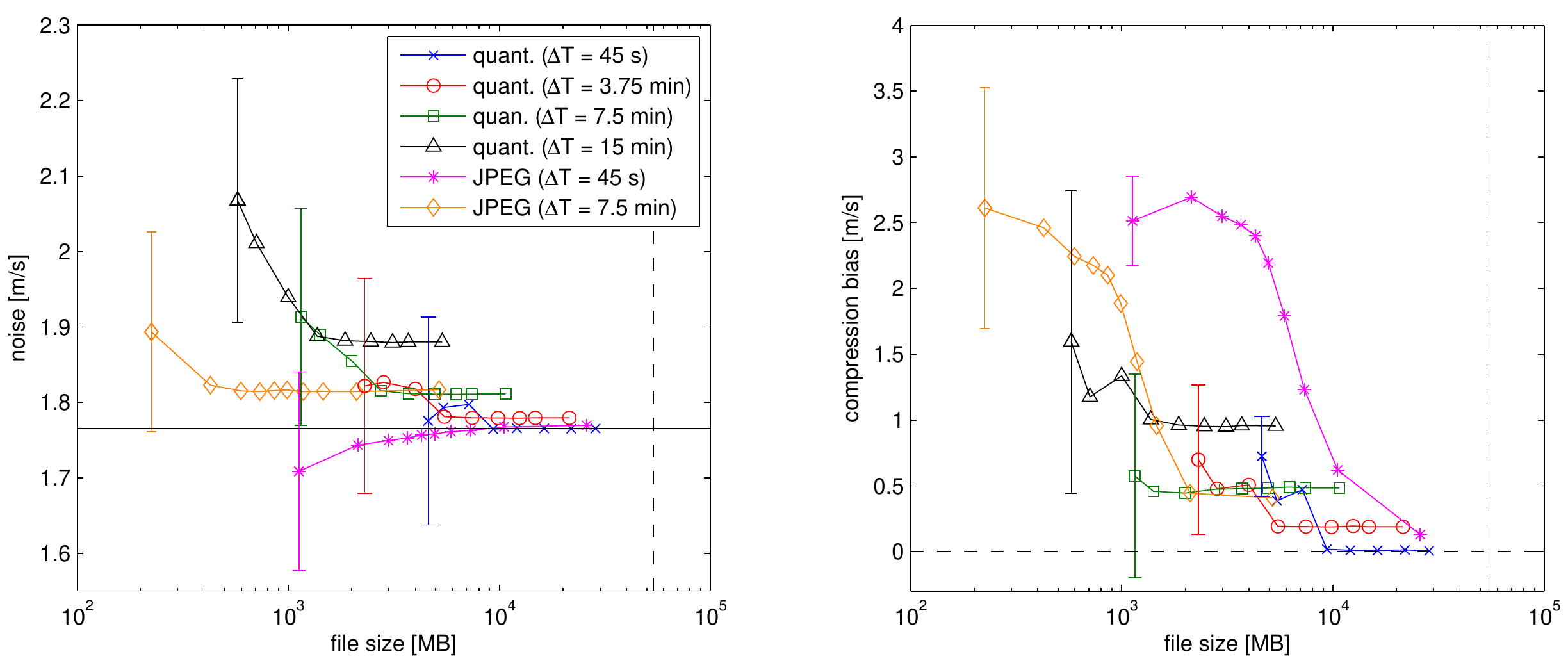}
\caption{Noise level ({\it left}) and absolute value of the compression bias ({\it right}) averaged over latitude (not including the data centered at $10^\circ$ latitude) as a function of file size for different compression methods. The different curves correspond to the different compression methods (quantization and JPEG) and to data with different time lags $\Delta t$ between consecutive pairs of images. {\it Blue crosses:} quantization with $\Delta T = 45$~s, {\it red circle:} quantization with $\Delta T = 3.75$~min, {\it green squares:} quantization with $\Delta T = 7.5$~min, {\it black triangles:} quantization with $\Delta T = 15$~min, {\it purple asterisks:} JPEG compression with $\Delta T = 45$~s, and {\it orange diamonds:} JPEG compression with $\Delta T = 7.5$~min. Along the curves, we vary the number $n$ of possible values for the intensity for quantization or the quality factor $q$ for JPEG compression. The file size represents the file size necessary for measuring differential rotation as in 
Figure~\ref{fig:rotation} ($T = 120$~days). The horizontal line in the left panel represents the noise level averaged over latitude for uncompressed data ($T = 120$~days). The vertical line shows the size of the uncompressed data (assuming a size of $7.1$~bits per pixel for the HMI raw data). For clarity, we only show one error bar for each curve. We only show results for short time lags. Larger time lags cause significantly higher compression bias and noise level.}
%\label{fig:power_spectra}
\label{fig:comp_size}
\end{figure*}

The amount of additional noise and the compression bias depend on the method and the parameters of the compression. Figure~\ref{fig:comp_size} shows compression bias and noise level averaged over latitude for all values of $n$ and $q$ and several time lags $\Delta T$. Quantization causes a much smaller compression bias than JPEG. Only very low values of $n$ lead to a significant bias. Both quantization and JPEG compression have almost no impact on the noise level. These methods introduce noise on small spatial scales, which is not relevant when measuring rotation. The time lag method increases the noise level and bias significantly.

The apodization parameter $\sigma$ has only little influence on the derived rotation velocities. The rotation velocities for $\sigma = 6$~pixels are almost indistinguishable from $\sigma = 10$~pixels, both for compressed and uncompressed data. The apodization affects mostly high wavenumbers, which are not relevant for measuring rotation. However, in case of $\sigma = 3$~pixels, there is some bias relative to the other values of $\sigma$ that we have tested, especially at high latitudes. This is caused by a significantly reduced sensitivity of LCT to flows for low values of $\sigma$.

\section{Discussion and conclusion}
The results of this paper suggest that LCT is robust to data compression when using quantization. However, we only looked at differential rotation and supergranulation, which does not need to be representative for LCT in general. Measurements that rely more on small spatial scales might be affected by an increased noise level to a higher degree. Also, we tested the influence of compression using tracked and remapped intensity images, applying it on raw full disk images might give different results. 

Figure~\ref{fig:comp_size} shows that the file size can be reduced by a factor of ten relative to the size of the HMI raw images without significant bias ($< 0.5$~m/s) or increased noise (no increase within the error bars). The influence of the compression bias relative to the noise level depends on the observing time. The expectation value of the bias is constant with time but the noise level decreases with increasing observing time. Both for individual flow maps and for differential rotation, quantization is the best compression method. JPEG compression leads to a higher noise level and it decreases the sensitivity of LCT more than quantization. JPEG compression removes information about small spatial scales in the continuum images, including the granulation signal used by LCT~\citep[see][]{2014A&A...571A..42L}. Quantized images on the other hand clearly show granulation, even for small {\it n}. \citet{2014A&A...571A..42L} found that JPEG was the best compression method tested for time-distance 
helioseismology.
This suggests applying different compression methods to the Dopplergrams used by local helioseismology and the continuum images used by LCT. The time lag method is very sensitive to changes of the geometry of the data. It can only be used if the pointing and roll angle of the instrument are both extremely stable with time. When deciding to use this method, it depends on the constraints of the observations which time lag $\Delta T$ to select. If both telemetry and observing time are fixed, the time lag should be as short as the telemetry allows, because this decreases the granulation noise more efficiently. If there is only a limit on telemetry and the image geometry is well known, observations with a large time lag for a long observing time are preferable. Suppose, the allocated telemetry is sufficient for transmitting a fixed number of continuum intensity images. In that case, obtaining these images over a long observing time with a high time lag $\Delta T$ between the individual flow maps is 
preferable because the granulation noise is highly correlated on small time scales. Sampling the granulation noise with a high cadence is not very efficient in decreasing the noise level. On the other hand, a longer observing time allows to cover a larger number of granules, leading to a lower noise level.

It might be possible to remove the compression bias from the flow velocities if its origin were understood. At least part of it could be caused by the shrinking-Sun effect. This effect and its coupling with compression might be responsible for the large differences between the northern and southern hemisphere in our rotation velocities. Our results motivate a detailed study of the origin of the shrinking-Sun effect and its observed asymmetry along the equator (see Figure~\ref{fig:flow_maps_quiver}). One possibility would be generating synthetic data from simulations of solar surface convection. Understanding the shrinking-Sun effect is also a requirement for applying LCT to {\it Solar Orbiter} data. Probing flows near the poles, even with {\it Solar Orbiter}, means observing far from disk center, where systematic effects become more important. In addition, the sensitivity of other feature tracking algorithms like the coherent structure tracking (CST) code~\citep{2007A&A...471..687R,2007A&A...471..695T,2013A&A...552A.113R} regarding systematic errors and compression needs to be tested.

In principle, it is also possible to run an LCT algorithm onboard the spacecraft and to transfer only the derived flow maps. This would decrease the required telemetry significantly, but would require an extremely good calibration of the instrument. The geometry of the data would have to be well known. Any systematic errors present in the continuum intensity maps would also affect the LCT flow maps.

A detailed study of the influence of data compression on LCT is not only important for {\it Solar Orbiter}, but also for other planned missions, such as L5 or SAFARI~\citep[see][for a review on concepts for future missions]{2015SSRv..tmp...15S}.

\begin{acknowledgements}
We are grateful to K. Nagashima for providing comparative data on differential rotation. We acknowledge support from Deutsche Forschungsgemeinschaft (DFG) through SFB 963/1 "Astrophysical Flow Instabilities and Turbulence" (Project A1). Support was also provided by European Union FP7 projects SPACEINN and SOLARNET. The German Data Center for SDO, funded by the German Aerospace Center (DLR), provided the IT infrastructure for this project. L.G. acknowledges support from the Center for Space Science, NYU Abu Dhabi Institute, UAE.
\end{acknowledgements}

\bibliographystyle{aa} % style aa.bst
\bibliography{literature} % your references Yourfile.bib

\end{document}